\documentclass{appolb}
\usepackage{graphicx}

\begin{document}
\title{ Role of the spectator system in electromagnetic effects
\thanks{Presented at the XIII Workshop on Particle Correlations and Femtoscopy, Krak\'ow, Poland, May 22–-26, 2018}%
}
\author{K. Mazurek$^1$,   A.~Rybicki$^1$, A. Szczurek$^1$, V.~Ozvenchuk$^1$, P.N. Nadtochy$^2$, 
A. Kelic$^3$, A. Marcinek$^1$ 
\address{$^1$Institute of Nuclear Physics Polish Academy of Sciences, PL-31342 Krakow, Poland\\
           $^2$Omsk State Technical University, Mira prospekt 11, Omsk, 644050, Russia\\
          $^3$GSI, Planckstrasse 1, 64291 Darmstadt, Germany
}
           }
\maketitle
\begin{abstract}
The electromagnetic effects on charged pion  ($\pi^+,\pi^-$) spectra provide new, independent information on the space-time evolution of the ultrarelativistic heavy ion collision.
The spectator life time and its 
excitation energy may also be of importance for the understanding 
of the space-time evolution of the participant zone. This paper gives an overview of our coordinated effort to understand the interplay between electromagnetic phenomena and processes related to the fragmentation of the spectator system at forward rapidity in peripheral $Pb$+$Pb$ collisions at top CERN SPS energies. Our study includes on one hand the experimental analysis of electromagnetic effects and corresponding phenomenological Monte Carlo simulations, and on the other hand dedicated theoretical calculations based on the Abrasion-Ablation model ABRABLA and the 4D Langevin approach.
\end{abstract}

\section{Introduction}

The two preceding contributions to this Workshop reported studies of spectator-induced electromagnetic (EM) effects in different nuclear collision systems measured at the CERN SPS and RHIC~\cite{sputowska:2018,kielbowicz:2018}. 
%
It is commonly accepted that the quark-gluon plasma is created in the participant zone and subsequently 
evolves to the hadronic phase, and that the two nuclear remnants,
spectators, fly away from the collision.
While the name ``spectators'' suggests that they do not 
take part in the reaction,
experimental evidence has been recalled in Ref.~\cite{sputowska:2018} that they induce an electromagnetic 
distortion on charged pion ($\pi^+,\pi^-$) spectra (see also~\cite{ar,rybicki:2007,rybicki:2011}).
While at high 
collision energies 
the $\pi^+$ and $\pi^-$ 
total multiplicities should be comparable, it appears that
 depending on the pion transverse 
momentum their distributions differ
significantly, as it 
is presented in Fig.~\ref{fig_01}. This phenomenon is caused by
electromagnetic interactions between charged pions and spectators.
In non-central collisions the electromagnetic fields modify trajectories 
of final state charged particles. Positively charged spectators repulse 
the $\pi^+$ mesons and attract the $\pi^-$ mesons. 
This behavior has been observed in the NA49 experiment in 
$Pb$+$Pb$ collisions at 158 GeV/nucleon~\cite{sputowska:2018}. At this Workshop it
was confirmed, for the first time, for the $Ar$+$Sc$ reaction at 150$A$~GeV/$c$
by the NA61/SHINE Collaboration~\cite{kielbowicz:2018}.
The aim of this paper is to complete the picture presented in~\cite{sputowska:2018,kielbowicz:2018} by considering the fate of the spectator system. This work will concentrate on peripheral $Pb$+$Pb$ collisions at the top SPS energy of 158 GeV/nucleon ($\sqrt{s_{NN}}=17.3$~GeV).

\section{EM effects in peripheral $Pb$+$Pb$ collisions}  
\label{em}

\begin{figure}[t]
\centering
\resizebox{0.87\textwidth}{!}{%
         \hspace*{0.1cm}
	 \includegraphics[angle=0,scale=0.17]{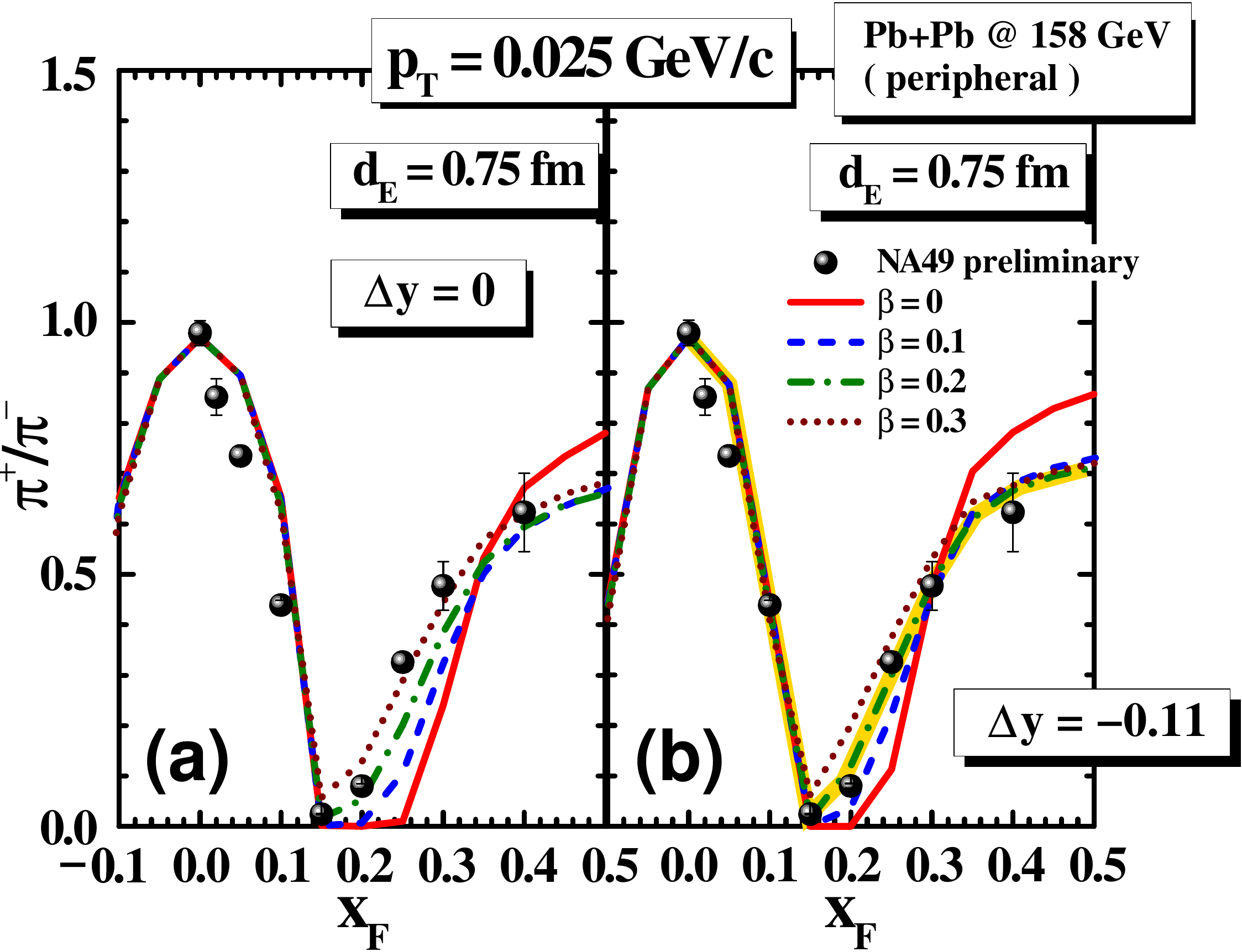} }
\caption{The charged pion ratio $\pi^+/\pi^-$ measured at transverse momentum $p_T$=0.025~GeV/$c$ in peripheral $Pb$+$Pb$ collisions at 150$A$ GeV. The experimental data from the NA49 experiment~\cite{ar} are compared to model simulations
obtained for different combinations of the parameters $d_E$, $\beta$, and $\Delta y$ described in the text.
  The optimal description of experimental~data is indicated by the yellow line. }
\label{fig_01}%
\vspace*{-0.2cm}
 \end{figure}

The information on the distance $d_E$ between the pion emission zone 
and the spectator was obtained with the help of
Monte Carlo simulations by a direct comparison to experimental data
for $\pi^+$ and $\pi^-$ production.
The best reproduction of experimental data for $Pb$+$Pb$ 
collisions at 158 GeV/nucleon was obtained for $d_E$=0.75~fm~\cite{sputowska:2018}.  
The new calculations, presented in Fig.~\ref{fig_01}, take into account also the
possibility of a radial expansion of the spectator with 
surface velocity $\beta$ treated here as a free parameter, and the possible change of spectator rapidity $\Delta y$. The result of the new calculation is consistent with the earlier value of $d_E$=0.75~fm, but we
obtain a better quantitative description of the NA49 data. As apparent from Fig.~\ref{fig_01}(a), a stable spectator ($\beta=0$) moving at the original beam velocity does not provide the optimal description of the data points, and the agreement is improved by assuming a considerable expansion velocity. We note that the optimal description is 
achieved once a shift $\Delta y=-0.11$ from the original beam rapidity is assumed in the model for the spherically expanding charged cloud, see Fig.~\ref{fig_01}(b). This we presently interpret as an indication for the presence of not only the spectator charge, but also the faster part of participant charge in the total charged cloud responsible for the EM effect.
Qualitatively, the presence of participant charge at high rapidity in peripheral $Pb$+$Pb$ collisions would be naturally expected from the energy-conservation-based model discussed at this Workshop~\cite{wpcf}. However, it is 
clear that the EM effect will be also sensitive to the expansion/fragmentation of the spectator system. For this reason, our studies of the latter will be summarized below.


\section{Simulation of the spectator system}

Basic kinematical considerations
suggest that
fast
pions 
at $p_T=0$
at top SPS energy can spend a time of the order of 400-2000 fm/$c$ in collision c.m.s.~in 
the
close 
vicinity
of the spectator system. This implies that we are mostly interested in the fate of the spectator remnant for the first few hundreds of fm/$c$ in its own c.m.s. 
Thus, a discussion of the spectator excitation energy and its 
de-excitation processes is necessary. This is studied within 
the abrasion-ablation statistical model (ABRABLA) 
\cite{gaimard:1991,kelic:2008}. This approach was employed 
previously with success in the description of relativistic collisions. 
Its application to ultrarelativistic energies, that is, to $Pb$+$Pb$ reactions at 158 GeV/nucleon considered here, was
presented in Ref.~\cite{mazurek:2018}. 
 
Information on the spectator evolution in time is extracted from models 
used in low-energy physics, based on solving transport equations 
of Langevin type. The dynamics of the spectator de-excitation is treated
as fission or particle and $\gamma$ evaporation processes. The ensemble 
of spectators created with the abrasion process by ABRABLA is evolved 
in space and time, by changing spectator shape, excitation energy, emitting 
particles and $\gamma$-rays. In the final stage, information on mass,
charge of the fission fragments and evaporation residua, as well as 
multiplicity and energy spectra of emitted particles and further 
observables is available. The details of the model and 
references can be found in Ref.~\cite{mazurek:2017}. The hybrid 
method of combining the abrasion part from the ABRABLA code with 
the dynamically solved set of Langevin equations brings the 
information on the space-time evolution of the spectator system. 

\section{Results}  

\begin{figure}
\centering
\resizebox{0.8\textwidth}{!}{%
	 \includegraphics[angle=0,scale=0.4]{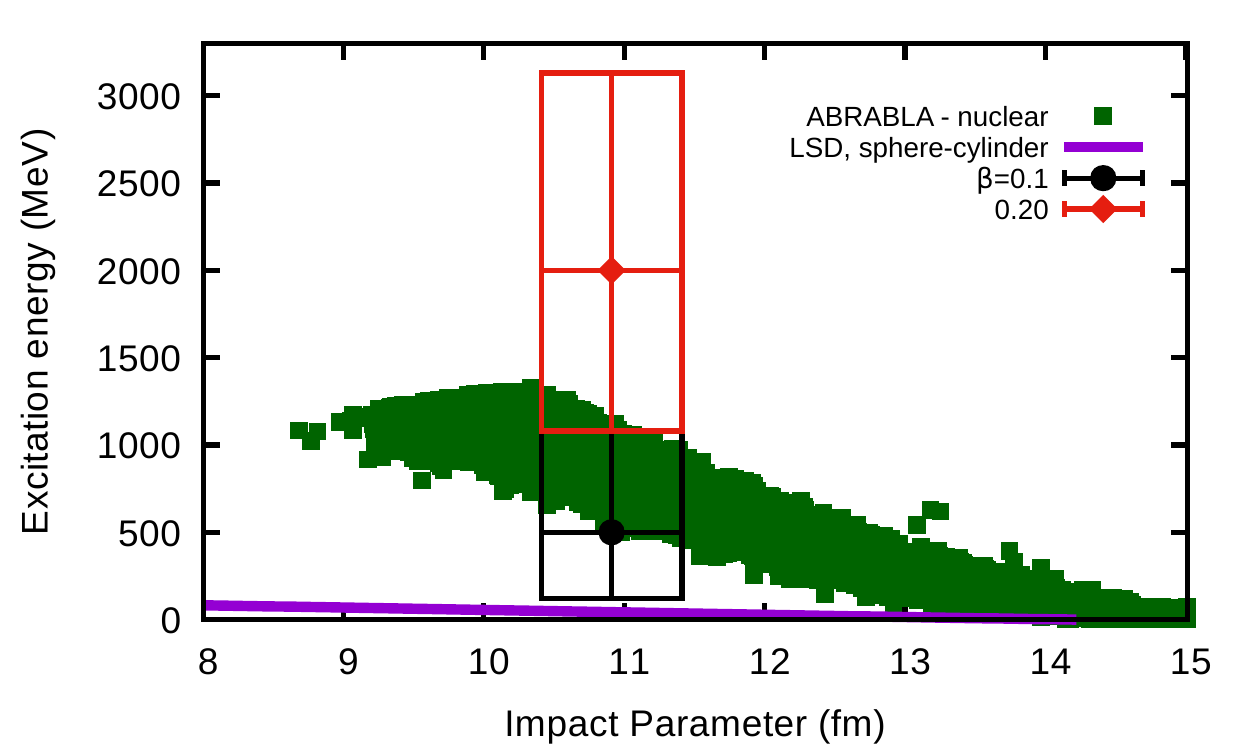} }
\caption{The distribution of the excitation energy of
  the spectator in $^{208}Pb$+$^{208}Pb$ collisions at 158 GeV/nucleon beam energy, predicted by the Abrasion-Ablation model ABRABLA (green squares), and
  assuming the geometrical model of sphere-cylinder collision (purple
  line)~\cite{mazurek:2018}. The red rectangle gives the estimated kinetic energy range for the expanding charged sphere corresponding to the optimal description of EM effects ($\beta=0.2\pm$0.05, yellow line in Fig.~\ref{fig_01}(b)). For comparison, the same estimate for $\beta=0.1\pm$0.05 is also presented.}
\label{fig_02}%
 \end{figure}

In Ref.~\cite{mazurek:2018} the predictions for excitation energies were
presented within different theoretical approaches.
For instance, the geometrical (macroscopic) approach was based on 
the assumption that the remnant of the collision of two nuclei 
at ultrarelativistic energies has a shape of a sphere cut by 
a cylinder. The deformation energy of such a shape was translated directly 
into the excitation energy of the spectator
within  
the Lublin-Strasbourg Drop approach, based on the liquid drop
model. This brought the deformation (excitation) energy of the order of 
100~MeV, indicated by the straight line in Fig.~\ref{fig_02}. 

On the other hand, the ABRABLA code estimates the excitation energy 
from a microscopic picture. Its estimation for the spectator excitation 
energy is displayed in Fig.~\ref{fig_02} (green squares). For comparison, the figure also includes the estimation of kinetic energy of an expanding charged sphere which gives the best description of NA49 data on electromagnetic effects ($\beta$=0.2$\pm$0.05). This is shown as the red box in Fig.~\ref{fig_02}, placed at the average impact parameter estimated for this data sample~\cite{b-k-g}. It is clear from section~\ref{em} that it is premature to make strong conclusions from this comparison: the presence of participant in addition to spectator charge will contribute to the expansion of the total charge cloud ``seen'' by pions at high rapidity. This will come on top of the evolution of the spectator {\it per se}. It is nevertheless evident that the order of magnitude of kinetic energy corresponding to the charge cloud apparent from EM effects is quite comparable to that of excitation energy obtained from ABRABLA.

\begin{figure}
\centering
\resizebox{0.8\textwidth}{!}{%
	 \includegraphics[angle=0,scale=0.4]{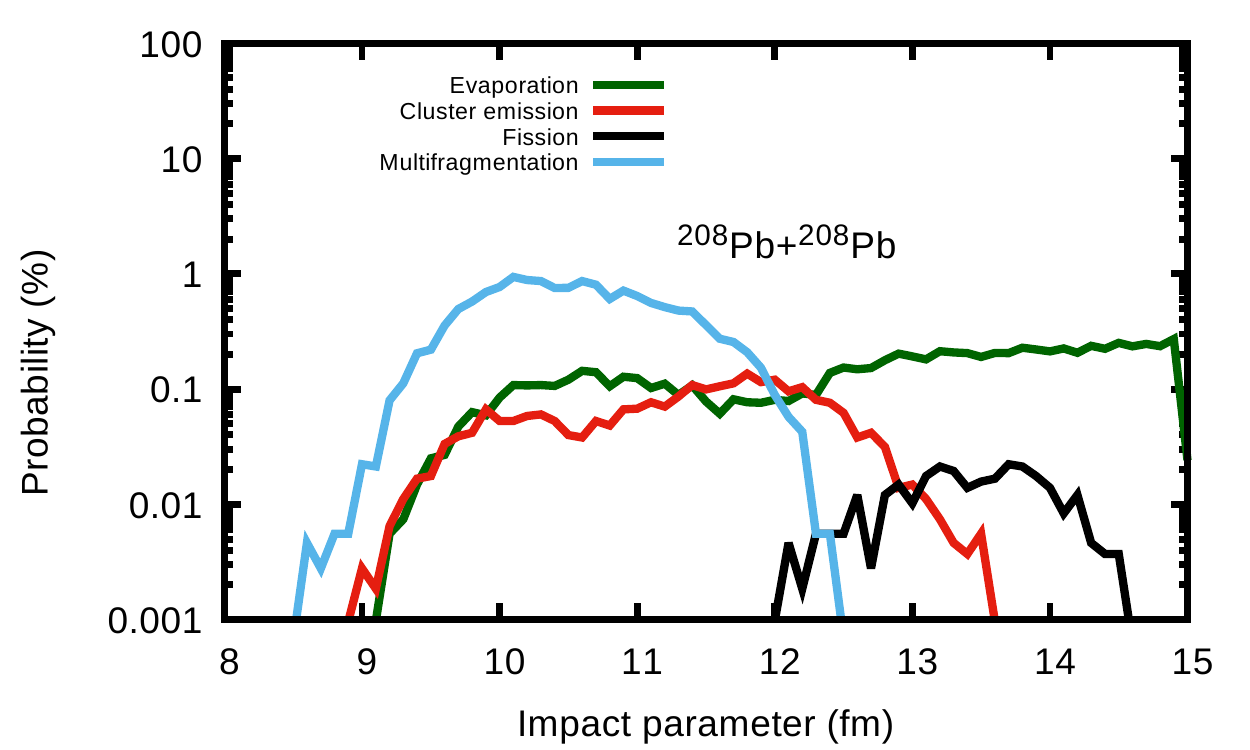} }
 \caption{The ABRABLA estimate of the probability of various nuclear processes (evaporation, fission, cluster emission and multifragmentation) as de-excitation channels for the hot spectator in $Pb$+$Pb$ collisions at 158 GeV/nucleon.}
\label{fig_03}%
\end{figure}

The subsequent fate of the spectator system can be illustrated by
probability of its various decay channels. These 
are
extracted from ABRABLA calculations incorporating statistical 
de-excitation. The result is presented in Fig.~\ref{fig_03}. 
The hot spectator can de-excite via nuclear processes
such as: evaporation, fission, cluster emission or multifragmentation. 
The vaporization and break-up reactions are omitted in this discussion;
they lead to almost immediate disintegration of the nucleus. 
Fission is possible only for peripheral collisions and 
multifragmentation dominates for smaller impact parameters. 
The evaporation of neutrons, protons and other light particles 
is almost independent of $b$ in the interval $b$=9$-$15 fm. 

In Ref.~\cite{mazurek:2018} we discussed the fission life time, but the latter fission
constitutes only a small part of the hot spectator de-excitation in Fig.~\ref{fig_03}. Therefore 
the evaporation of particles: neutrons, protons and $\gamma$-rays 
may possibly give a better estimation for typical life times of the spectator system. 
During the cooling down of the spectator, the Langevin code provides 
information on the evaporation time of each particle. As it appears from the calculations, neutron 
emission is much more probable than other emissions.
For example for the impact parameter $b$=10.5~fm, 
the mean multiplicity of emitted neutrons is 26, while that of protons is~6.
 \begin{figure}
\centering
\resizebox{0.8\textwidth}{!}{%
	 \includegraphics[angle=0,scale=0.40]{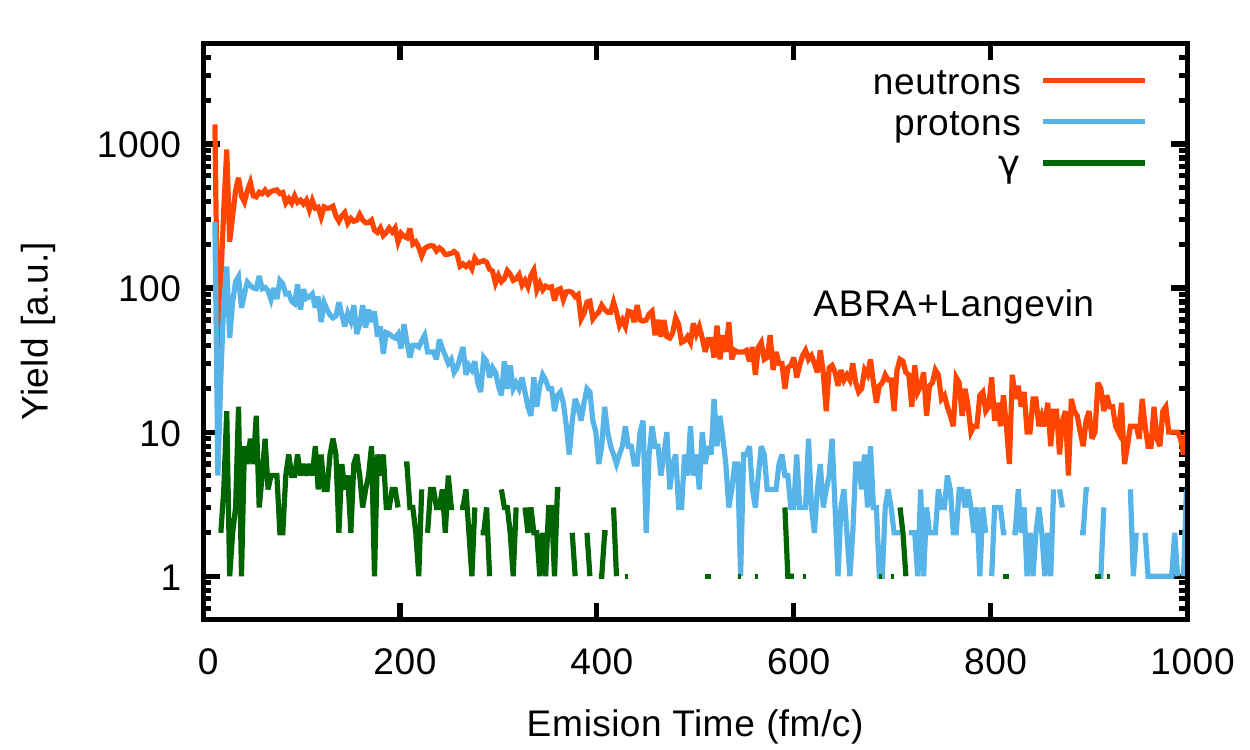} }
  \caption{The time distribution (in spectator c.m.s.) of neutron, proton
    and $\gamma$ emission in the evaporation process. The initial ensemble of spectators
    has been generated by the ABRABLA code and de-excited via 
the 4DLangevin code.}
\label{fig_04}%
\end{figure}

Fig.~\ref{fig_04} shows the distribution of emission time of 
evaporated
neutrons, protons, and~$\gamma$. The corresponding calculation was performed with 
the 4DLangevin code, where the initial conditions were taken from 
ABRABLA, similarly as in Ref.~\cite{mazurek:2018}. 
Our
4DLangevin prediction 
shows that
nucleons are mostly evaporated in a time below 600 fm/$c$ in the spectator c.m.s.
Account taken of the Lorentz boost ($\gamma\approx 9.2$ at this collision energy),  appears quite comparable to our estimated time of 400--2000 fm/$c$ for the charged pion to remain in close spectator vicinity. This implies that evaporation and the EM effect will interplay in the course of the $Pb$+$Pb$ collision.


 
\section{Summary}


Spectator-induced EM effects on charged particle distributions 
provide new information on the space-time evolution of 
the system created in nucleus-nucleus collisions. 
Up to now, modeling of these effects in heavy ion collisions suffered from lack of 
knowledge on the time evolution of the spectator system. On the other hand, 
it was known that this evolution was 
an important issue in the corresponding phenomenological studies. 
A first coordinated effort has been undertaken to investigate 
this problem from both sides (experimental data on EM effects and 
phenomenological simulations, {\it versus} dedicated nuclear theory). 
First results are encouraging: for peripheral $Pb$+$Pb$ collisions considered here, multifragmentation seems to be the dominant spectator de-excitation channel.
 It seems indicated to obtain more information on this latter channel in our future studies.
 Fission seems to play a more significant role only for
much more peripheral $Pb$+$Pb$ collisions, but neutron, proton and $\gamma$ evaporation has non-negligible probability for a wide range of impact parameters. 
The spectator excitation energy obtained from theoretical calculations, and our first phenomenological estimates of kinetic energy of expansion of the effective charged cloud, have begun to converge. Nonetheless, first indication 
for the presence of participant on top of spectator charge in this latter cloud is apparent from EM effects, which could provide another verification of the collective model based on energy-momentum conservation~\cite{wpcf}. It is 
our hope that these inter-disciplinary studies can help us to improve our
understanding of both the longitudinal evolution of the QGP and 
the excitation and decay of the spectator system.


\vskip 1cm
{\bf Acknowledgements}
\\
The work was partially supported by the Polish
National Science Centre under contract no.~2013/08/M/ST2/00257 (LEA COPIGAL) (project no.~18) and under grant no.~2014/14/E/ST2/00018.


\end{document}